\documentclass[fleqn,twoside]{article}
\usepackage{espcrc2}

\newcommand{\AmS}{{\protect\the\textfont2
  A\kern-.1667em\lower.5ex\hbox{M}\kern-.125emS}}

\title{Physics development of web-based tools for use in hardware 
	clusters doing lattice physics}

\author{P. Dreher\address[MIT]{MIT Laboratory for Nuclear Science, 
        Massachusetts Institute of Technology,\\ 
        Cambridge, MA  02139}%
        \thanks{Support provided through the U.S. Department of Energy 
Cooperative
     Agreement DE-FC02-94ER40818}%
        \thanks{Member, Lattice Hadron Physics Collaboration}
	\thanks{Poster presenter},
W. Akers\address[JLAB]{Thomas Jefferson National Accelerator Facility,\\
                     Newport News, VA  23606}\thanks{This work was supported by the 
U.S. Department of Energy contract DE-AC05-84ER40150, under which the
Southeastern Universites Research Association (SURA) operates the Thomas
Jefferson National  Accelerator Facility},
J. Chen\addressmark[JLAB]$^\S$,
Y. Chen\addressmark[JLAB]$^\S$,
C. Watson\addressmark[JLAB]$^{\dagger\S}$}
\begin{document}

\begin{abstract}
Abstract:

Jefferson Lab and MIT are developing a set of web-based tools within the Lattice Hadron 
Physics Collaboration to allow lattice QCD theorists to treat the computational facilities located 
at the two sites as a single meta-facility. The prototype Lattice Portal provides researchers the 
ability to submit jobs to the cluster, browse data caches, and transfer files between cache and 
off-line storage. The user can view the configuration of the PBS servers and to monitor both the 
status of all batch queues as well as the jobs in each queue. Work is starting on expanding the 
present system to include job submissions at the meta-facility level (shared queue), as well as 
multi-site file transfers and enhanced policy-based data management capabilities.

\vspace{1pc}
\end{abstract}

\maketitle

\section{THE META-FACILITY CONCEPT}

The next generation of computers for lattice calculations and other grand challenge type numerical 
simulation problems will be constructed with multi-teraflops computing capabilities.  Such machines
dedicated to a particular grand challenge problem may not be geographically located at a single site.
The Thomas Jefferson National Accelerator Facility and the MIT Laboratory for Nuclear Science have 
undertaken a joint project to develop a meta-facility for lattice physics calculations at a 
geographically dispersed facility.  

A ``meta-facility" for lattice physics will consist of a set of tools and utilities that allow for the efficient 
distribution of of data and computing tasks among multiple lattice gauge facilities. 
The goals of the meta-facility will provide a central 
location for users to submit and monitor these jobs, collect and distribute data generated from
these computing activities, and status of the overall status of the 
hardware interconnected to the meta-facility.

The full implementation of the meta-facility will have the capability of providing a primary 
routing mechanism to provide users with the ability to run jobs at one of several sites 
according to the resources currently available at any given time.  At each site, there will 
be a batch system that accepts jobs from the central routing queue.  
A distributed batch system at each site will queue the job(s) sent to each site
and control the execution of the jobs on the machines at that site. 
The distributed batch system at each site will also communicate with the central routing queue 
providing job status and control information to the routing queue.

\section{THE PRESENT JLAB-MIT FACILITY}

At the present time, JLab and MIT have developed an initial set of web-based tools and utilities 
that have been installed at both sites.  This initial software deployment will become part of the 
design toward an eventual full implementation of a meta-facility at both sites.  

These web based tools have installed to monitor the clusters of machines at both JLab and MIT.  
These machines currently have a mix of different capabilities and configurations.  

Jefferson Lab presently has two clusters operational.  The first cluster consists of 
16 XP1000 nodes from Compaq.   
The XP1000 is a 64 bit Alphaa 21264 processor (500 MHz)with a 4 MB L2 cache and 64 KB on chip cache, 
a 100 MHz SDRAM subsystem and integrated 4 GB 
Wide-Ultra SCSI (10000rpm) disk subsystem with dual independent 
32/64 bit PCI buses.
Eight of the XP1000 nodes are have Myrinet cards installed,  THe nodes are connected to a 
Myrinet switch.
The second cluster consists of 12 dual processor UP2000 from API Labs.
The UP-2000 systems have dual 667 MHz 21264 processors with 4 MB caches,
512 MB memory, and an 18 GBIDE disk (7200 rpm)

At MIT, the are twelve Compaq ES40 machines.  Each ES40 machine has four EV-67 667 Mhz 
alpha processors and 1 Gbyte of memory in an SMP configuration within each box.  The twelve 
ES40s are connected by both Myrinet hardware and fast ethernet.  There are also additional
Intel based PCs that serve as backup, file and batch servers at this site.  This entire set of 
machines is positioned behind an Intel based front-end machine.  This front end machine serves 
the functionality of a firewall and as a repository for the web-based tools and utilities 

The current capabilities of the present system allow users at either the JLab or MIT sites 
to launch and monitor batch jobs and examine batch queue parameters and machine status.
The JLab site also has the capability to allow users to browse the tape file catalog and 
initiate file transfers from to and from the JLab tape storage facilities.
At the present time there is a certificate server operational at JLab with the capability 
of issuing a personal web certificate for enhanced security for users submitting jobs 
to that site.

\section{FUTURE PLANS AND ENHANCEMENTS}

The present configuration that is operational at JLab and MIT 
has demonstrated a proof of concept.  The batch system
execution queues at both the JLab and MIT sites operate properly and a Web based 
link has been established between both sites that provides information on the status 
and batch jobs running on all clusters.
The project will be moving forward with two major upgrade tasks scheduled over the 
next several months.

- At the present time, batch job submission via the Web is limited to the JLab site.  
Web services for batch submission of jobs at the MIT site will be installed in the Fall 
of 2001.

- Tape transfers at both JLab and MIT involve a two-stage process because of the front-end
machines acting as firewalls at both sites.  Work is ongoing to develop full tape file 
transfer capability at both sites 

After these immediate tasks have been completed and are operational, there are several 
longer term activities planned for development and implementation of a fully functional
meta-facility.

At both the MIT and JLab sites, only the execution queues have been configured in the 
batch queuing system at each site.  The ultimate goal is to develop and install a 
separate node in the meta-facility that will be dedicated to the functionality of 
a routing queue.  The routing queue would not be attached to any of the individual 
clusters in the meta-facility but would rather view all of the execution queues on 
every cluster at all sites throughout the meta-facility.  The routing node will act 
as a separate communications and traffic manager.  

Users would only 
submit jobs to the node that was running the routing queue system software.  The routing 
queue would have the configuration parameters of each execution queue and communicate 
with all of the various execution queues located throughout the entire system.  Based on
the job requirements, the routing queue would examine
systems loads and available resources throughout the entire meta-facility and direct the 
job to the execution queue on the cluster with the most available resources.
At this point additional sites beyond the Jefferson Lab and MIT locations would be added.

Finally, the routing queue will be integrated with future data grid software currently being 
developed using Java and XML-based web services \cite{Watson}.  The final system will be 
a web-based distributed data grid that is site independent.  It will include a distributed 
batch system augmented with various monitoring tools and management options to deliver a 
full production level meta-facility for lattice physics.

\end{document}